# Understanding inelastically scattered neutrons from water on a time-of-flight small-angle neutron scattering (SANS) instrument

Changwoo Do[a*], William T. Heller[a], Christopher Stanley[a], Franz X[b]. Gallmeier, Mathieu Doucet[c], and Gregory S. Smith[a]

[a]Biology and Soft Matter Division, Neutron Sciences Directorate, Oak Ridge National Laboratory, Oak Ridge, TN 37831, USA
[b]Instrument & Source Design Division, Neutron Sciences Directorate, Oak Ridge National Laboratory, Oak Ridge, TN 37831, USA
[c]Neutron Data Analysis & Visualization Division, Neutron Sciences Directorate, Oak Ridge National Laboratory, Oak Ridge, TN 37831, USA

[*]To whom correspondence should be addressed; E-mail: doc1@ornl.gov.

**ABSTRACT**

It is generally assumed by most of the small-angle neutron scattering (SANS) user community that a neutron's energy is unchanged during SANS measurements. Here, the scattering from water, specifically light water, was measured on the EQ-SANS instrument, a time-of-flight (TOF) SANS instrument located at the Spallation Neutron Source of Oak Ridge National Laboratory. A significant inelastic process was observed in the TOF spectra of neutrons scattered from water. Analysis of the TOF spectra from the sample showed that the scattered neutrons have energies consistent with room-temperature thermal energies (~20 meV) regardless of the incident neutron's energy. With the aid of Monte Carlo particle transport simulations, we conclude that the thermalization process within the sample results in faster neutrons that arrive at the detector earlier than expected based on the incident neutron energies. This thermalization process impacts the measured SANS intensities in a manner that will ultimately be sample- and temperature-dependent, necessitating careful processing of the raw data into the SANS cross-section.

# 1. Introduction

Small-angle neutron scattering (SANS) is a powerful tool for investigating the structure of materials at the nanoscale that is applicable to materials of interest to condensed matter physics, the materials sciences, chemistry, and the biosciences. SANS measurements are performed using instruments installed at both steady-state and time-of-flight (TOF) pulsed neutron sources. The basic physical design of the instruments at the two types of sources are essentially the same, but the scattered intensity as a function of momentum transfer ($q$) is collected using a wide band of neutron wavelengths at a pulsed source, while SANS instruments at steady state sources use a nearly monochromatic band of neutron wavelengths. In both cases, the $q$-range, statistics, and the resolution can be varied by the selection of a combination of wavelength band, sample to detector distance, and instrument optics. In general, TOF-SANS instruments have the advantage of capturing a larger $q$-range in a single instrument setting compared to steady-state SANS instruments, which generally have the ability to probe smaller values of $q$ as a result of a higher time-averaged flux of the long wavelength neutrons that are better suited to low-$q$ measurements.

SANS data are typically analyzed assuming that the absolute cross-section produced by the data reduction process results strictly from elastic scattering [1]. However, several scattering processes take place when a neutron interacts with a sample, including absorption, incoherent scattering and inelastic scattering. Of these effects, inelastic scattering is generally assumed to be negligible in SANS experiments. However, this assumption is often inaccurate in the case of samples containing a large amount of hydrogen, which constitutes many samples studied by SANS as a result of the ability to substitute deuterium for hydrogen and dramatically alter the scattering from a sample. Water, organic solvents and polymeric materials, all of which contain a great deal of hydrogen, produce strong isotropic scattering that results from the large incoherent scattering cross section of hydrogen. Additionally, inelastic-incoherent effects resulting from hydrogen comprise a significant portion of the sample-dependent background. These effects are known and methods have been developed to account for these processes when using water as a secondary calibration standard [1-2]. These inelastic scattering processes can also impact data correction operations, including background subtraction and sample transmission corrections, or when using hydrogenous materials as calibration standards [2-4]. Here, we present experimental data from the EQ-SANS at the Spallation Neutron Source of Oak Ridge National Laboratory[5], a TOF-SANS, that demonstrates how a considerable fraction of the scattered neutrons undergo a significant energy change after interacting with a water sample. Through the use of advanced Monte Carlo simulations on the interaction of neutrons with a water sample, it is found that the scattering processes lead to a thermalization of the scattered neutrons.

## 2. Materials and methods

*2.1. Time-of-Flight SANS*

Small-angle neutron scattering (SANS) measurements were performed at the EQ-SANS instrument at the Spallation Neutron Source (SNS) located at Oak Ridge National Laboratory (ORNL) using 60 Hz operation. A sample-to-detector distance of 4 m was used with various wavelength bands collectively covering the $q$ range of 0.004 Å$^{-1}$ < $q$ < 0.3 Å$^{-1}$, where $q = (4\pi/\lambda) \sin(\theta/2)$ is the magnitude of the scattering vector, and $\theta$ is the scattering angle. The instrument was configured to provide a 1 Å-wide incident bandwidth by appropriately setting the phases of the four disk choppers [5]. Data on the EQ-SANS instrument are collected and saved in 'event mode' where the time-of-flight to the detector and the detector pixel position is recorded for each detected neutron. The SNS source operates at 60 Hz, providing a 16,667 μs measurement time window at the detector. Data processing was accomplished with the MantidPlot software package (http://www.mantidproject.org). The light water sample, consisting of deionized water from a Barnstead water purification system, was loaded into a standard 1 mm path length amorphous quartz cell (Hellma USA). Graphite foil, which produces little incoherent scattering, served as a reference elastic scattering sample.

*2.2. Simulation methods*

Simulations of the scattering of neutrons by a thin light water sample were performed with the Monte Carlo particle transport code MCNPX version 2.6.0 [6]. In addition to describing the transport of a multitude of particle types with energies ranging from $10^{-5}$ eV to 100 GeV through complex three-dimensional structures, the MCNPX code can simulate neutron scattering in condensed matter by utilizing so-called material-specific scattering kernels and includes that of light water [7]. For this study, the scattering kernels for light water were prepared from ENDF/B-VII evaluation files [7] into ACE-formatted scattering kernels in the energy-probability table format that can be read by the software. The scattering formalism based on the energy-probability table format is better-suited for simulating scattering of mono-directional neutron beams compared to the discrete-energy table format.

The interaction of mono-directional neutron beams having 1 Å wide wavelength bands between 4 and 11 Å with a 1 mm thick water sample were simulated. The simulated, scattered neutrons were collected in a square detector of 1 $m^2$ area positioned 4 meters away from the sample to match the configuration of the EQ-SANS instrument employed. The neutron spectra were accumulated into energy bands ranging from $10^{-5}$ eV to 1 eV that were divided into 20 bins per decade of energy. The 30 mm diameter area around the incident beam direction corresponding to the transmitted beam was not included in the analysis. The number of scattering events that the neutron experienced in the sample was retained for use in the simulation analysis.

3. Results

In **Figure 1**, the transmitted empty beam spectrum of a 4-5Å wavelength band measured at the EQ-SANS detector is shown within the 16,667 μs neutron pulse period, termed a 'frame', and multiple frames are shown to more easily visualize the data as it is collected during an experiment. It is important to note that the time shown in these plots is not the absolute travel time of the neutron from the sample to detector, but rather the relative arrival time in relation to the burst of neutrons from the target. At 4m sample to detector distance of EQ-SANS, approximately 4-7 Å neutrons arrive at the detector within the second frame (16,667 μs < $T_{TOF}$ < 33,334 μs) after they are produced in the target and 8-11 Å neutrons arrive within the third frame (33,334 μs < $T_{TOF}$ < 50,001 μs) in terms of the flight time from the source ($T_{TOF}$). The flux at EQ-SANS peaks near 2.5 Å and decreases almost monotonically with increasing wavelength [5], giving rise to the spectral shape. The slope between 1000 and 5000 μs shows this behaviour. The sharp spikes at or near 0 and 16665 μs are detection events resulting from the $T_0$ pulse from the source that are not filtered by either the bent guide of the instrument or by the current chopper settings of measurements. The transmitted neutron spectrum of water ($H_2O$) in the quartz cell (**Figure 1a** blue) is essentially identical to the empty beam spectrum (**Figure 1a** red) when scaled by the transmission coefficient of $H_2O$. However, the scattered spectrum from water obtained by summing over all detector pixels outside of the direct beam (**Figure 1b** green) has a very different shape than the TOF spectrum resulting from both the scattered and transmitted spectra of the empty cell. The spectrum shows a much steeper slope between 1000 and 5000 μs, and a considerable number of neutrons are detected near and before $t = 0$ μs, indicating that neutrons are arriving faster than the incoming neutrons (i.e. empty beam, **Figure 1b** red). The earlier arrival time necessitates that the neutrons gain energy through their interaction with the water sample. Here, the fraction of neutrons that gained energy is estimated to be 15% of total number of neutrons produced by that particular source pulse. The TOF spectrum from the quartz cell, measured for the same number of monitor counts, was observed to be very weak. (**Figure 1b** blue) In **Figure 1c**, the scattered flux spectrum was summed over ring-shaped regions about the detector center. Each ring has a 5 cm width and the inner radii of the rings are 5, 15, 25, 35, and 45 cm, which covers $0.0157 \text{ Å}^{-1} < q < 0.195 \text{ Å}^{-1}$ overall with wavelengths of 4 to 5 Å. Each spectrum is normalized to a maximum of 1 to enable clear comparison of the spectral shapes. The curves overlap almost perfectly, indicating that the change in the scattered neutron energies resulting in the fast-arriving neutrons near $t = 0$ μs is independent of $q$. The results demonstrate that a non-negligible number of inelastic scattering processes take place in SANS measurements of water. Vibrational modes of water molecules are known to exist at much higher $q$ values [8-9], but not in the $q$ range of the measurements presented here.

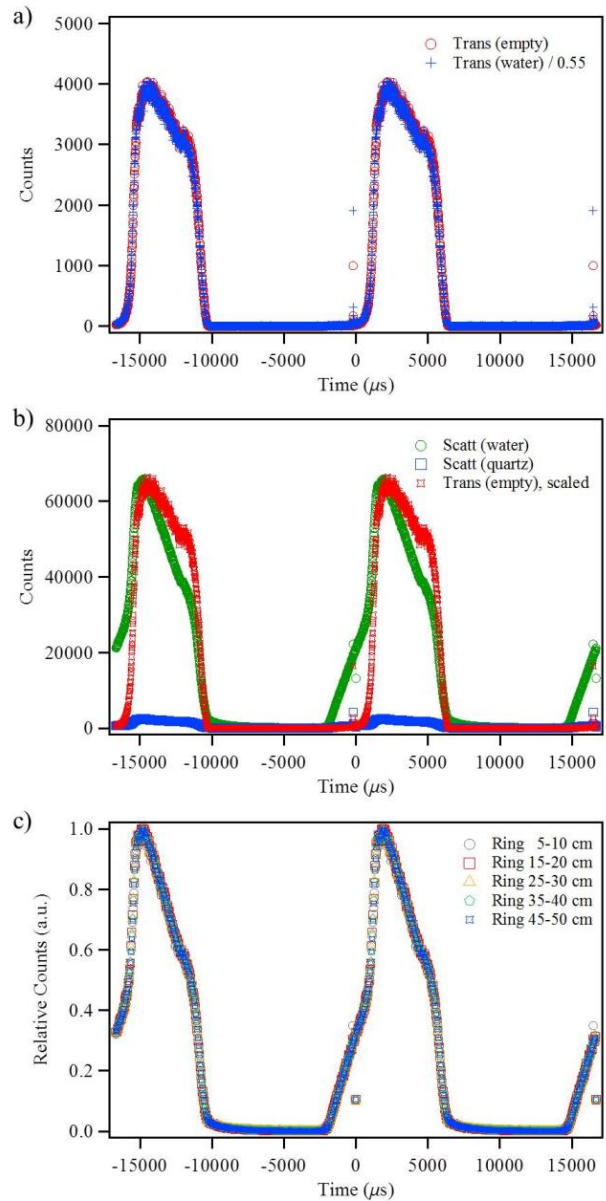

**Figure 1.** a) TOF spectrum of the empty beam (red) and the water +quartz cell transmitted beam (blue) of 4-5A wavelength band. The transmitted beam from water was scaled to allow easy comparison with the spectrum of empty beam. b) The TOF spectrum of the scattered beam from (water + quartz cell) (green), the quartz cell only (blue) and the empty beam spectrum (red). The empty beam spectrum was scaled for ready comparison with the scattering spectrum of the water. c) The TOF spectrum from water at the detector with different scattering angles. Each spectrum has been normalized to 1 for clear comparison.

To verify that the observed change in the scattered neutron spectrum results from neutrons experiencing a change in energy, measurements of graphite foil were performed. The scattering from graphite foil is purely elastic and the material is regularly used as an instrument resolution standard for neutron spin-echo (NSE) instruments [10], because it does not exhibit any dynamics under the spatial and energy ranges of NSE. To maximize the effect of any energy change experienced by a scattered neutron, neutrons with long wavelength (12 Å), and thus relatively low energy (ca. 0.57 meV) were used with the full wavelength band possible for this configuration of the EQ-SANS (3.5 Å). Here, the time axis represents the detector arrival time within the third frame and the start of the wavelength band is located at approximate 5000 μs. (**Figure 2**) The sharp edges of the TOF spectra are clearly observed at both $t \sim -12000$ μs and $t \sim 5000$ μs, and are identical to those of the direct beam. The results indicate that no fast-arriving neutrons have resulted after scattering by the graphite foil sample. Comparison of the scattered neutron spectra of water and graphite foil makes it possible to conclude that the neutron arrival time changes result from changes in energy resulting from scattering by the sample.

To better characterize and understand the origins of the energy shift resulting from scattering by water, a series of TOF-SANS experiments were performed on the same water sample using a series of different incident neutron wavelengths with the instrument choppers phased to provide a 1 Å wavelength band. Neutrons having wavelengths of 6, 8, and 10 Å, correspond to incident energies of 2.25 meV, 1.27 meV, and 0.81 meV, respectively. The scattered neutron spectra are presented in **Figure 3.** In all cases, neutrons can be seen at to the left of the incident neutron TOF spectrum (i.e. apparently before the pulse starts) regardless of the incident neutron's energy. The TOF spectral shapes of these early-arriving neutrons differs from the incident neutron spectrum, and so the energy difference cannot be defined precisely. However, the beginning of each wavelength band is at the peak of the intensity profiles of incident neutrons, so it can be assumed that neutrons arriving between $t_{min}$ and $t_{max}$, as defined in **Figure 3**, that comprise the peak prior to the incident pulse have been accelerated from the 6, 8, and 10 Å neutrons, respectively, by energies having a well-defined distribution. From this assumption, the accelerated time of arrival can be estimated to be $\Delta t \sim$ 4025, 6125, and 8150 $\mu$s for neutrons having incident wavelengths of 6, 8, and 10 Å, respectively. Neutrons having 6, 8, and 10 Å wavelengths require $t_6 \sim$ 6067 $\mu$s, $t_8 \sim$ 8089 $\mu$s, and $t_{10} \sim$ 10111 $\mu$s to travel the 4 m sample-to-detector distance. Considering the time gains experienced by the up-scattered neutrons ($\Delta t$), the flight time for the neutrons of **Figure 3a-c** can be calculated to be $t_{accel,6} \sim$ 2042 $\mu$s, $t_{accel,8} \sim$ 1964 $\mu$s, and $t_{accel,10} \sim$ 1961 $\mu$s. In terms of velocity of neutrons, these correspond to 1959, 2037, and 2040 m/s, which again lead to 20.3, 21.9, and 22.0 meV, respectively. One can immediately notice that these values are very close to the energy scale of room temperature thermal energy, $k_BT \sim$ 25.9 meV at 300K, where $k_B$ is the Boltzmann constant and T is the temperature. Therefore, we speculate that these neutrons which form the peak of early arriving neutron in the TOF spectra are neutrons that have been thermalized as a result of scattering by water, causing them to arrive at the detector earlier than the incident neutrons. The thermalization process is governed by the temperature of water, so it is expected that the energy of a thermalized neutron would be independent of the incoming neutron energy and the change in energy would be greater for longer wavelength (lower energy) neutrons, as observed. The portion of neutrons being accelerated through interaction with the sample increases with increasing wavelength (**Figures 3a-c**). In the case of 10-11 Å neutrons (**Figure 3c**), nearly 60 % of the scattered neutrons are thermalized.

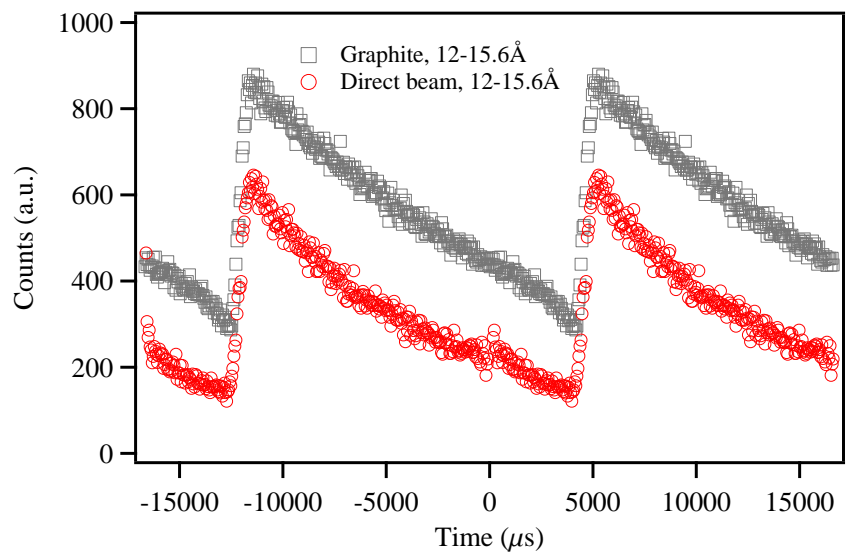

**Figure 2.** TOF spectrum of empty beam (red) and scattered beam from graphite foil (grey) of 12-15.6 Å wavelength band. The graphite scattering TOF spectrum was scaled arbitrarily to help visual comparison.

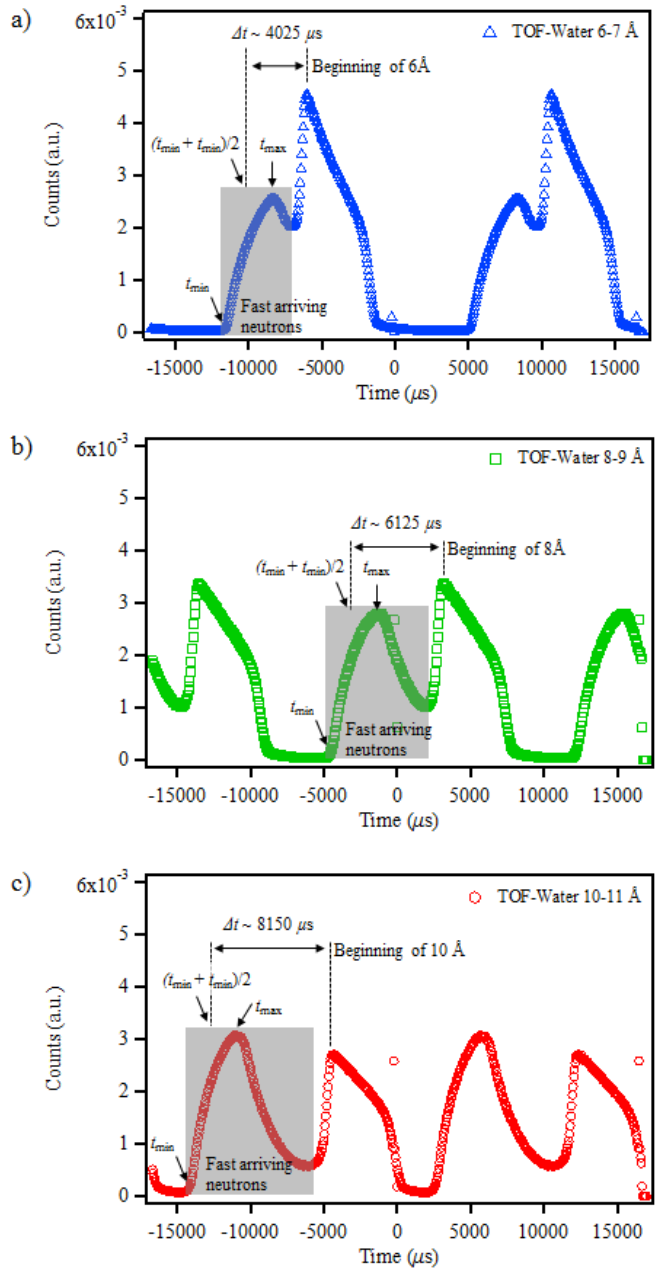

**Figure 3.** TOF spectrum of the scattered beam from water with different incident wavelength bands: (a) 6-7 Å, (b) 8-9 Å, and (c) 10-11 Å. Fast-arriving neutrons are indicated with grey shading.

Thermalization is a well-known process from nuclear reactor design. Additionally, most neutron scattering facilities use water and other hydrogenous materials as moderators for producing thermal or cold neutrons with different energy spectra. To understand the changes in the neutron spectra resulting from scattering by the water sample, general purpose Monte-Carlo radiation transport code, MCNPX, was employed to simulate the scattering process. The simulations were configured to replicate the experimental conditions, having a sample-to-detector of 4 m and an incident neutron wavelength bandwidth of 1 Å. As observed in the TOF-SANS experiments with 6-7 Å, 8-9 Å, and 10-11 Å wavelength bands, simulations of neutron scattering by 1 mm of water resulted in a large number of thermalized neutrons having a short wavelength of 1~3 Å range (2Å ~ 20 meV), as can be seen in the scattered neutron spectra presented in **Figure 4**. The relative ratio between the thermalized neutrons to the neutrons with initial energy also increases with increasing wavelength (lower energy of incident neutrons), which is consistent with the experimental results presented in **Figure 3**. The scattered neutrons in the simulation were further categorized depending on the number of scattering events that took place to understand the influence of multiple scattering on the energy changes observed. (**Figure 5**) The results clearly demonstrate that neutrons undergoing two or more scattering events have gained energy and reached thermal equilibrium with the water. (**Figure 5**)

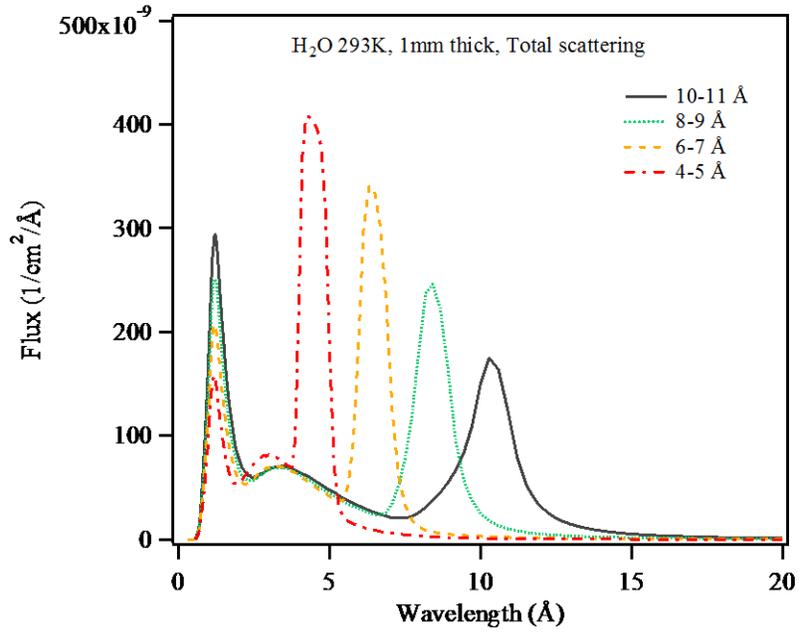

**Figure 4.** Flux spectrum of simulated scattered neutrons from water having different incident wavelength bands.

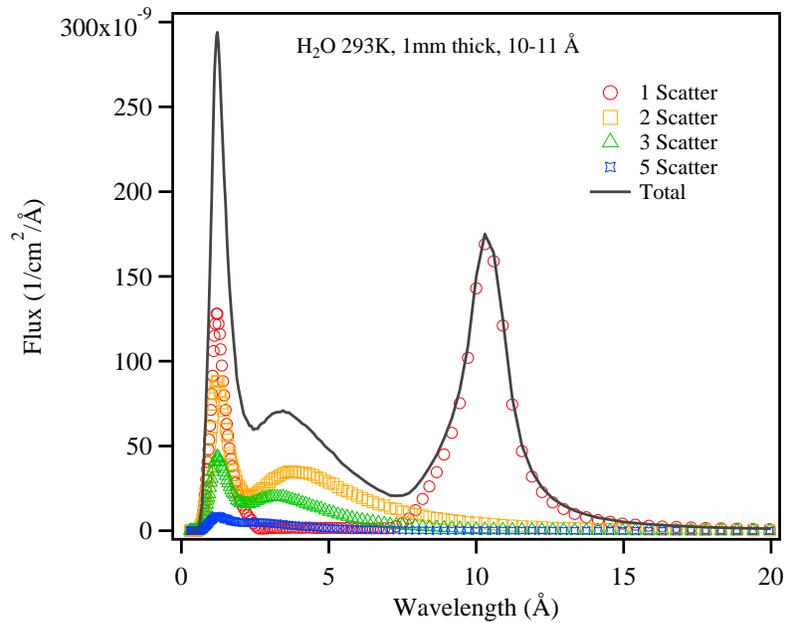

**Figure 5.** Flux spectrum of simulated scattered neutrons from water with 10-11 Å neutrons. The fluxes of neutrons with different number of scattering events are summarized. The solid line is for the total flux spectrum from all the scattering events.

## 4. Conclusions

Changes in the neutron spectra resulting from scattering by water as measured on a TOF-SANS instrument have been investigated. Unlike the generally-accepted idea that a neutron's energy remains unchanged during small-angle scattering measurements, the results demonstrate that significant inelastic processes take place as a result of scattering by water that manifest in changes in the TOF spectra from that of the incident neutron spectra. TOF analysis showed that the scattered neutrons have energies consistent with room temperature thermal energies (~20 meV) regardless of the incident neutron's energy. With the aid of MCNPX simulations, it is clear that thermalization is responsible for the faster neutrons produced by scattering in the sample. The faster neutrons manifest in a TOF-SANS measurement as neutrons arrive at the detector earlier than expected based on the incident neutron spectrum. The influence of thermalization by water is more significant for longer wavelength incident neutrons and the simulation shows that initial neutron energy spectrum has been mostly lost after two scattering events in the sample.

In TOF-SANS, the wavelength assigned to a neutron, and thus the $q$-value calculated, depends on the time-of-arrival of neutrons at the detector. Therefore, any neutrons that are inelastically scattered in the sample complicate the wavelength-dependent normalization that takes place during data reduction. The thermalization of neutron energies primarily results from incoherent scattering and multiply scattered neutrons are almost entirely thermalized. This suggests that avoiding multiple scattering is even more critical in TOF-SANS experiments in order to minimize the thermalization effect. These observations also suggest that careful background subtraction and sample preparation are very critical for performing successful TOF-SANS experiments. The effect can be reduced by performing measurements at shorter wavelengths and by performing very careful background subtraction rather than relying on a flat incoherent scattering background, as is generally applicable to SANS data collected at steady-state sources.

**Acknowledgements**

This Research at Oak Ridge National Laboratory's Spallation Neutron Source was sponsored by the Scientific User Facilities Division, Office of Basic Energy Sciences, U.S. Department of Energy.